\documentclass[]{aastex701}

\begin{document}

\title{Multi-height Identification of Sausage and Fluting Eigenmodes in a Solar Pore}

\author[orcid=0000-0002-7711-5397]{Shahin Jafarzadeh}
\affiliation{Astrophysics Research Centre, School of Mathematics and Physics, Queen’s University Belfast, Belfast, BT7 1NN, UK}
\affiliation{Max Planck Institute for Solar System Research, Justus-von-Liebig-Weg 3, 37077 G\"{o}ttingen, Germany}
\email[show]{shahin.jafarzadeh@qub.ac.uk}  

\author[orcid=0000-0002-9155-8039]{David B. Jess} 
\affiliation{Astrophysics Research Centre, School of Mathematics and Physics, Queen’s University Belfast, Belfast, BT7 1NN, UK}
\affiliation{Department of Physics and Astronomy, California State University Northridge, Northridge, CA 91330, USA}
\email[]{}  

\author[orcid=0000-0002-5365-7546]{Marco Stangalini} 
\affiliation{ASI Italian Space Agency, Via del Politecnico snc, I-00133 Rome, Italy}
\email[]{}   

\author[orcid=0000-0002-5082-1398]{Luiz A. C. A. Schiavo} 
\affiliation{Northumbria University, Newcastle upon Tyne NE1 8ST, UK}
\email[]{}

\author[orcid=0000-0003-3306-4978]{Timothy J. Duckenfield}
\affiliation{Astrophysics Research Centre, School of Mathematics and Physics, Queen’s University Belfast, Belfast, BT7 1NN, UK}
\email[]{} 

\author[orcid=0000-0001-5414-0197]{Suzana S. A. Silva} 
\affiliation{Plasma Dynamics Group, School of Electrical and Electronic Engineering, The University of Sheffield, Sheffield, UK}
\email[]{}

\author[orcid=0000-0002-9546-2368]{Gary Verth} 
\affiliation{Plasma Dynamics Group, School of Mathematical and Physical Sciences, The University of Sheffield, Sheffield, UK}
\email[]{}

\author[orcid=0000-0002-0893-7346]{Viktor Fedun} 
\affiliation{Plasma Dynamics Group, School of Electrical and Electronic Engineering, The University of Sheffield, Sheffield, UK}
\email[]{}

\author[orcid=0000-0002-3418-8449,sname='Solanki']{Sami~K.~Solanki} \affiliation{Max Planck Institute for Solar System Research, Justus-von-Liebig-Weg 3, 37077 G\"{o}ttingen, Germany}\email{solanki@mps.mpg.de}	

\author[orcid=0000-0003-3490-6532,sname='Smitha']{H.~N.~Smitha} \affiliation{Max Planck Institute for Solar System Research, Justus-von-Liebig-Weg 3, 37077 G\"{o}ttingen, Germany}\email{narayanamurthy@mps.mpg.de}	

\author[orcid=0000-0003-1459-7074,sname='Lagg']{Andreas~Lagg} \affiliation{Max Planck Institute for Solar System Research, Justus-von-Liebig-Weg 3, 37077 G\"{o}ttingen, Germany}\email{lagg@mps.mpg.de}	

\author[orcid=0000-0002-9972-9840,sname='Gandorfer']{Achim~Gandorfer} \affiliation{Max Planck Institute for Solar System Research, Justus-von-Liebig-Weg 3, 37077 G\"{o}ttingen, Germany}\email{gandorfer@mps.mpg.de}

\author[orcid=0009-0009-4425-599X,sname='Feller']{Alex~Feller} \affiliation{Max Planck Institute for Solar System Research, Justus-von-Liebig-Weg 3, 37077 G\"{o}ttingen, Germany}\email{feller@mps.mpg.de}	

\author[orcid=0000-0003-1409-1145,sname='Iglesias']{Francisco~A.~Iglesias} \affiliation{Max Planck Institute for Solar System Research, Justus-von-Liebig-Weg 3, 37077 G\"{o}ttingen, Germany}\affiliation{Grupo de Estudios en Heliofísica de Mendoza, CONICET, Universidad de Mendoza, Boulogne sur Mer 683, 5500 Mendoza, Argentina}\email{iglesias@mps.mpg.de}	

\author[orcid=0000-0001-6317-4380,sname='Riethmüller']{Tino~L.~Riethmüller} \affiliation{Max Planck Institute for Solar System Research, Justus-von-Liebig-Weg 3, 37077 G\"{o}ttingen, Germany}\email{riethmueller@mps.mpg.de}	

\author[sname='Grauf']{Bianca~Grauf} \affiliation{Max Planck Institute for Solar System Research, Justus-von-Liebig-Weg 3, 37077 G\"{o}ttingen, Germany}\email{grauf@mps.mpg.de}	

\author[orcid=0000-0001-6029-7529,sname='Hoelken']{Johannes~Hoelken} \affiliation{Max Planck Institute for Solar System Research, Justus-von-Liebig-Weg 3, 37077 G\"{o}ttingen, Germany}\email{hoelken@mps.mpg.de}
\author[orcid=0000-0002-5054-8782,sname='Katsukawa']{Yukio~Katsukawa} \affiliation{National Astronomical Observatory of Japan, 2-21-1 Osawa, Mitaka, Tokyo 181-8588, Japan}\affiliation{Department of Astronomy, The University of Tokyo, 7-3-1, Hongo, Bunkyo-ku, Tokyo 113-0033, Japan}\affiliation{Department of Astronomical Science, The Graduate University for Advanced Studies (SOKENDAI), 2-21-1 Osawa, Mitaka, Tokyo 1818588, Japan}\email{yukio.katsukawa@nao.ac.jp}	
\author[orcid=0000-0002-0787-8954,sname='Bernasconi']{Pietro~Bernasconi} \affiliation{Johns Hopkins University Applied Physics Laboratory, 11100 Johns Hopkins Road, Laurel, Maryland, USA}\email{pietro.bernasconi@jhuapl.edu}	
\author[sname='Berkefeld']{Thomas~Berkefeld} \affiliation{Institut für Sonnenphysik (KIS), Georges-Köhler-Allee 401a, 79110 Freiburg, Germany}\email{thomas.berkefeld@leibniz-kis.de}	
		
\author[orcid=0000-0001-9228-3412,sname='Álvarez-Herrero']{Alberto~Álvarez-Herrero} \affiliation{Instituto Nacional de T\'ecnica Aeroespacial (INTA), Ctra. de Ajalvir, km. 4, E-28850 Torrejón de Ardoz, Spain}\affiliation{Spanish Space Solar Physics Consortium}\email{alvareza@inta.es}	
\author[orcid=0000-0001-5616-2808,sname='Kubo']{Masahito~Kubo} \affiliation{National Astronomical Observatory of Japan, 2-21-1 Osawa, Mitaka, Tokyo 181-8588, Japan}\email{masahito.kubo@nao.ac.jp}	
\author[orcid=0000-0001-8829-1938,sname='Orozco~Suárez']{David~Orozco~Suárez} \affiliation{Instituto de Astrofísica de Andalucía, CSIC, Glorieta de la Astronomía s/n, 18008 Granada, Spain}\affiliation{Spanish Space Solar Physics Consortium}\email{orozco@iaa.es}	
\author[sname='Carpenter']{Michael~Carpenter} \affiliation{Johns Hopkins University Applied Physics Laboratory, 11100 Johns Hopkins Road, Laurel, Maryland, USA}\email{michael.carpenter@jhuapl.edu}	
\author[sname='Bell']{Alexander~Bell} \affiliation{Institut für Sonnenphysik (KIS), Georges-Köhler-Allee 401a, 79110 Freiburg, Germany}\email{albe@leibniz-kis.de}	
\author[orcid=0000-0001-7764-6895,sname='Martínez~Pillet']{Valentín~Martínez~Pillet} \affiliation{Instituto de Astrofísica de Canarias, Vía Láctea, s/n, E-38205 La Laguna, Spain}\affiliation{Spanish Space Solar Physics Consortium}\email{vmpillet@iac.es}
		
\author[orcid=0000-0002-7318-3536,sname='Bailén']{Francisco~Javier~Bailén} \affiliation{Instituto de Astrofísica de Andalucía, CSIC, Glorieta de la Astronomía s/n, 18008 Granada, Spain}\affiliation{Spanish Space Solar Physics Consortium}\email{fbailen@iaa.es}	
\author[orcid=0000-0002-2055-441X,sname='Blanco~Rodríguez']{Julian~Blanco~Rodríguez} \affiliation{Universitat de Valencia Catedrático José Beltrán 2, E-46980 Paterna-Valencia, Spain}\affiliation{Spanish Space Solar Physics Consortium}\email{julian.blanco@uv.es}	
\author[orcid=0000-0003-4319-2009,sname='Castellanos~Durán']{Juan~Sebastián~Castellanos~Durán} \affiliation{Max Planck Institute for Solar System Research, Justus-von-Liebig-Weg 3, 37077 G\"{o}ttingen, Germany}\email{castellanos@mps.mpg.de}	
\author[orcid=0009-0002-6808-5154,sname='Harnes']{Edvarda~Harnes} \affiliation{Max Planck Institute for Solar System Research, Justus-von-Liebig-Weg 3, 37077 G\"{o}ttingen, Germany}\email{harnes@mps.mpg.de}	
\author[orcid=0000-0002-4669-5376,sname='Ishikawa']{Ryohtaroh~T.~Ishikawa} \affiliation{National Institute for Fusion Science, 322-6 Oroshi-cho, Toki City 509-5292, Japan}\email{ishikawa.ryohtaro@nifs.ac.jp}	
\author[orcid=0000-0001-7452-0656,sname='Kawabata']{Yusuke~Kawabata} \affiliation{National Astronomical Observatory of Japan, 2-21-1 Osawa, Mitaka, Tokyo 181-8588, Japan}\email{kawabata.yusuke@nao.ac.jp}	
\author[orcid=0000-0002-1043-9944,sname='Matsumoto']{Takuma~Matsumoto} \affiliation{Centre for Integrated Data Science, Institute for Space-Earth Environmental Research, Nagoya University, Furocho, Chikusa-ku, Nagoya, Aichi 464-8601, Japan}\email{takuma.matsumoto@gmail.com}	
\author[orcid=0000-0002-7044-6281,sname='Oba']{Takayoshi~Oba} \affiliation{Advanced Research Center for Space Science and Technology, Institute of Science and Engineering, Kanazawa University, Kakuma-machi, Kanazawa, Ishikawa 920-1192, Japan}\affiliation{Max Planck Institute for Solar System Research, Justus-von-Liebig-Weg 3, 37077 G\"{o}ttingen, Germany}\email{oba@mps.mpg.de}	
\author[orcid=0000-0003-0175-6232,sname='Siu-Tapia']{Azaymi~L.~Siu-Tapia} \affiliation{Instituto de Astrofísica de Andalucía, CSIC, Glorieta de la Astronomía s/n, 18008 Granada, Spain}\affiliation{Spanish Space Solar Physics Consortium}\email{siu@iaa.es}	
\author[orcid=0000-0003-1483-4535,sname='Strecker']{Hanna~Strecker} \affiliation{Instituto de Astrofísica de Andalucía, CSIC, Glorieta de la Astronomía s/n, 18008 Granada, Spain}\affiliation{Spanish Space Solar Physics Consortium}\email{streckerh@iaa.es}	
\author[orcid=0000-0003-1971-5551,sname='Vukadinović']{Dušan~Vukadinović} \affiliation{Institut für Physik, Universität Graz, Universitätsplatz 5, 8010 Graz, Austria}\affiliation{Max Planck Institute for Solar System Research, Justus-von-Liebig-Weg 3, 37077 G\"{o}ttingen, Germany}\email{vukadinovic@mps.mpg.de}	

\begin{abstract}
Magnetic pores are compact, strongly magnetised waveguides in the lower solar atmosphere and therefore provide favourable conditions for identifying magnetohydrodynamic (MHD) wave modes. Earlier seeing-free observations revealed concurrent sausage, kink, and fluting modes in photospheric pores, but only at a single sampled layer. In this Letter, we exploit the dense spectral sampling of the near-ultraviolet 327--329\,nm window observed by the {\sc Sunrise-iii} UV Spectropolarimeter and Imager (SUSI) to investigate how pore wave modes behave across multiple photospheric and low-chromospheric heights spanning roughly 500\,km. We analyse $\sim$75\,min of a {\sc Sunrise-iii}/SUSI time series containing a small solar pore. From eight selected spectral lines sampling different estimated formation heights, we identify the pore boundary at each line and time step and apply proper orthogonal decomposition (POD) to the boundary oscillations. In all eight lines, the first POD mode is consistently identified as an axisymmetric sausage mode, with dominant power at $\sim$1--2\,mHz, and carries the dominant normalised eigenvalue fraction, typically about 66--86\%, while the second mode is a fluting mode with azimuthal wave number $m=2$, dominant at $\sim$2--3.5\,mHz, and contributes about 4--10\%. Cross-line wavelet phase analysis of the temporal coefficients shows that the sausage mode remains close to zero phase difference across the sampled heights, consistent with standing or near-standing behaviour, whereas the fluting mode displays a modest but systematic increase in phase with height, reaching about $50^{\circ}$, indicative of an upward-propagating component. These observations provide the first multi-height identification and phase characterisation of sausage and fluting modes inferred from pore-boundary oscillations.
\end{abstract}

\section{Introduction}
\label{sec:introduction}

A central challenge in solar physics is to understand how energy is transported through the magnetised atmosphere and redistributed between its different layers. Magnetohydrodynamic (MHD) waves are widely regarded as one of the main agents in this process, since they can channel, store, convert, and dissipate energy in structured magnetic environments from the photosphere to the corona \citep[e.g.,][]{2005LRSP....2....3N, 2015SSRv..190..103J, 2015LRSP...12....6K}. Theory predicts that magnetic structures can support a rich spectrum of coexisting MHD wave modes, each with its own spatial morphology, propagation properties, and energetic implications \citep{1983SoPh...88..179E, 2019mwsa.book.....R}. Determining which modes are present, how they evolve with height, and whether they are standing, propagating, or evanescent, therefore remains a key step toward understanding the dynamics and energetics of the solar atmosphere \citep{2023LRSP...20....1J}.

Recent advances in high-resolution observations and analysis methods have revealed concurrent MHD wave modes in sunspots \citep{2017ApJ...842...59J, 2021RSPTA.37900181A, 2022NatCo..13..479S, 2022ApJ...927..201A, 2023ApJ...954...30A}, in both the photosphere and chromosphere, by decomposing oscillatory perturbations of the sunspot structure. Surface modes have also been inferred in sunspots from spectropolarimetric perturbations in the umbra-penumbra boundaries \citep{2018ApJ...869..110S}. In addition, longer-period oscillatory power, extending from about 10--15\,min to several tens of minutes, has also been reported in sunspots and related magnetic structures \citep[e.g.,][]{2008SoPh..248..395S, 2010A&A...513A..27C, 2013A&A...554A.146K}. Pores, however, offer a particularly useful laboratory for such studies. They are smaller and geometrically simpler than sunspots, yet remain strongly magnetised and spatially resolved, making them especially suitable for tracking boundary motions and identifying surface wave modes.

Previous pore studies have established that pores host a broad range of oscillatory behaviour, including compressive sausage modes identified through area, intensity, and Doppler-velocity fluctuations \citep[e.g.,][]{2011ApJ...729L..18M, 2015ApJ...802...45C, 2015ApJ...806..132G, 2016ApJ...817...44F, 2018ApJ...857...28K}, as well as both standing \citep{2014A&A...563A..12D} and upwardly propagating \citep{2015A&A...579A..73M, 2022ApJ...938..143G} sausage disturbances. Coexistence of several azimuthal harmonics, including $m=0$, $m=1$, and $m=2$, has also been reported in a pore \citep{2019ApJ...877L...9K}. More recently, a decomposition analysis of seeing-free observations from the High Resolution Telescope \citep[HRT;][]{2018SPIE10698E..4NG} of the Polarimetric and Helioseismic Imager \citep[PHI;][]{2020A&A...642A..11S} on board Solar Orbiter \citep[SO;][]{2020A&A...642A...1M} demonstrated that several eigenmodes may coexist in the same pore, including sausage, kink, and an $m=2$ fluting mode \citep{2024A&A...688A...2J}. That result, however, was restricted to a single well-sampled photospheric layer. It therefore remained unclear how pore boundary perturbations behave when the same structure is sampled through a sequence of closely spaced atmospheric layers and whether their vertical phase behaviour can distinguish standing from propagating character of oscillations. This missing vertical information is the key obstacle addressed here in our Letter. Previous mode detections in solar pores have either relied on single-height diagnostics or on limited height separation between observing channels. A dense multi-line spectroscopic sequence, by contrast, allows the same magnetic structure to be sampled through an ordered ensemble of atmospheric response functions, making it possible to ask whether the same boundary eigenmodes persist with height and whether their temporal coefficients remain phase-aligned or acquire a systematic phase lag.

The third flight of the {\sc Sunrise} balloon-borne solar observatory has now opened that opportunity \citep{solankietal2026}. Building on the success of the earlier {\sc Sunrise} flights \citep{2010ApJ...723L.127S, 2011SoPh..268....1B, 2017ApJS..229....2S}, {\sc Sunrise-iii} \citep{2025SoPh..300...75K} is not just stabilized by a new gondola \citep{bernasconietal2025} and correlating wavefront sensor \citep{berkefeldetal2026}, it also carries a new suite of instruments designed for seeing-free, high-resolution, multi-line observations of the lower solar atmosphere. In particular, the {\sc Sunrise} UV Spectropolarimeter and Imager (SUSI) provides slit-based spectroscopy and spectropolarimetry in the near-ultraviolet (near-UV), allowing the same magnetic structure to be sampled in several closely spaced spectral lines with different atmospheric sensitivities \citep{2025SoPh..300...65F, 2025SoPh..300...58I}. This makes SUSI especially well suited to investigating how pore oscillations appear through a sequence of closely spaced sampled heights.

In this Letter, we use {\sc Sunrise-iii}/SUSI observations in the 327--329\,nm window to investigate the boundary oscillations of a small pore across eight selected spectral lines spanning different estimated formation heights. We identify the pore boundary in line-core intensity rasters, apply proper orthogonal decomposition (POD; \citealt{lumley1967}) to the boundary oscillations, and then compare the temporal behaviour of the dominant modes between the selected lines. This provides a first multi-height view of the leading oscillatory modes inferred from pore-boundary oscillations in the near-UV and allows their vertical phase behaviour to be examined directly. The central result is that the dominant sausage mode remains nearly phase-aligned across the sampled layers, whereas the weaker $m=2$ fluting mode shows a systematic upward phase progression.

\section{Data}
\label{sec:data}

This Letter uses a subset of the {\sc Sunrise-iii}/SUSI observations obtained on 14 July 2024 during the main science sequence on a large active region near disc centre \citep[$\mu \approx 0.96$; {\sc Sunrise-iii} ID 24\_SPOT,][]{solankietal2026}. SUSI sampled the 327--329\,nm window in a repeated small-raster mode, producing a narrow two-dimensional scan that crossed a sunspot, surrounding plage, and a nearby pore under seeing-free conditions. The observations span 21:40--23:39\,UTC and comprise 180 raster scans with a cadence of about 39\,s. The analysis is based on the SUSI Stokes~$I$ data products, using the standard pipeline reduction \citep{solankietal2026} together with subsequent assembly of time-series raster images and additional correction for solar rotation, field derotation, and residual image jitter. The resulting data cube samples the active-region pore for the full duration of the observations.

The present analysis is restricted to the final 115 raster scans, corresponding to about 75\,min of data. The first 65 scans were excluded because flare-related activity in the pore during that interval produced strong transient brightenings, particularly in some of the higher-forming lines, compromising the stability and consistency of the pore-boundary identification. The retained part of the sequence is free from these strong early transients and therefore provides a more reliable basis for a consistent multi-line boundary analysis.

The 327--329\,nm spectral region contains a dense set of absorption lines from multiple atomic and molecular species. From this window, we restrict the present analysis to eight spectral lines selected on the basis of relatively high signal-to-noise ratio, limited blending, stable boundary detection, and broad coverage of estimated formation heights. The selected lines are Fe\,{\sc ii} 327.6617\,nm, Mn\,{\sc i} 327.8562\,nm, Ti\,{\sc ii} 327.2094\,nm, Fe\,{\sc i} 327.2606\,nm, Fe\,{\sc i} 327.8742\,nm, Ni\,{\sc i} 328.6960\,nm, Fe\,{\sc i} 328.6772\,nm, and Cu\,{\sc i} 327.3972\,nm. Their representative formation heights are adopted from the same modelling framework used in \citet[][hereafter Papers~I and II]{Jafarzadeh2026_paper1, Jafarzadeh2026_paper2} and are used here primarily as approximate proxies for relative ordering rather than as precise geometrical heights. In short, they were estimated at each spatial pixel of a 3D radiative-MHD MURaM sunspot simulation \citep[][see also \citealt{2023A&A...669A.144S}]{2005A&A...429..335V, 2012ApJ...750...62R} through contribution functions (CFs) to the emergent intensity, computed under local thermodynamic equilibrium (LTE) with the RH code \citep{2001ApJ...557..389U, 2015A&A...574A...3P}, using atomic line parameters from the Kurucz line list \citep{1995all..book.....K}. The weighted-mean CF heights and their 16th--84th percentile ranges are therefore used here as a practical summary of the vertical sensitivity of the selected lines. As a result, we do not interpret these CF centroids as sharply separated geometrical layers. Instead, the selected lines are treated as an ordered set of partly overlapping atmospheric CFs. The robust observable in the present analysis is therefore the systematic cross-line behaviour of the POD temporal coefficients, rather than an absolute geometrical phase speed.

Unlike Papers~I and II, where the emphasis was on fitted line-core intensity and LOS-velocity products derived from Voigt-profile analysis \citep{2026FrASS...Jafarzadeh}, the present study uses line-core intensity raster images to track the oscillatory behaviour of the pore boundary. For each selected line, we extracted two-dimensional raster images at the chosen line-core wavelength position and followed the pore boundary in time within those image sequences. We did not apply an explicit Doppler compensation of the line-core wavelength for this purpose. Since the goal here is boundary identification in intensity images, rather than a pixel-by-pixel measurement of line-centre shifts, small Doppler-induced displacements of the local absorption minimum are not expected to alter the overall pore outline materially in the selected lines. The resulting image sequences form the basis for the contour extraction and POD analysis described in Section~\ref{sec:analysis_results}.

\begin{figure*}
\centering
\includegraphics[width=\textwidth]{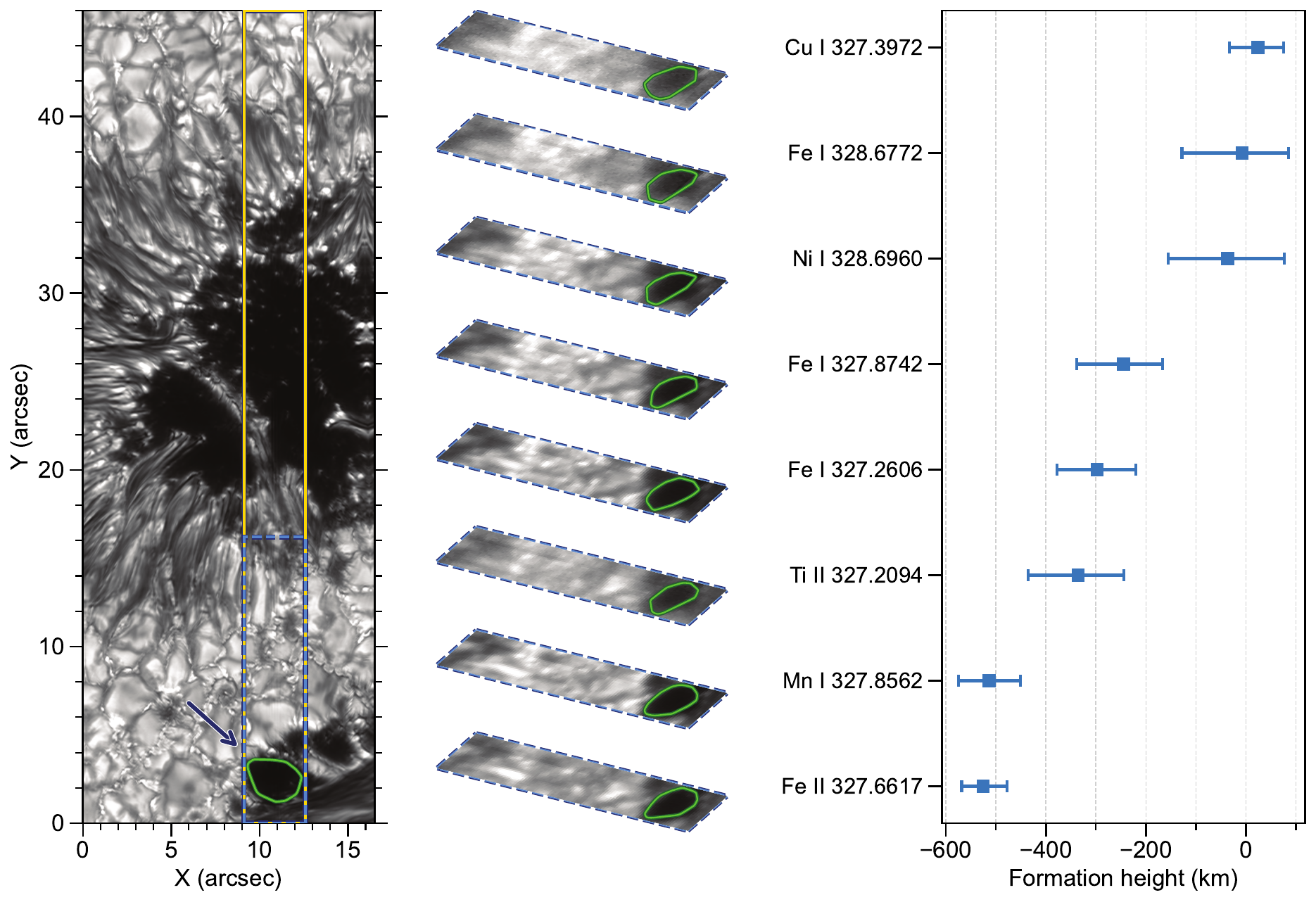}
\caption{Observational context and selected spectral lines used in the present analysis. The left panel shows a SUSI slit-jaw image near the continuum. The yellow lines mark the full raster field of view, while the dashed blue rectangle marks the smaller portion of that field shown in the middle panel. The arrow and green contour indicate the pore analysed in this work. The middle panel presents representative line-core intensity raster images for the eight selected spectral lines, displayed as a three-dimensional stack and ordered by increasing effective formation height. The dashed blue outlines mark the boundaries of the displayed raster cutouts, and the green contours indicate the pore boundaries overplotted at each sampled height. The right panel shows the adopted formation heights of the same lines. Symbols indicate the weighted-mean contribution-function heights, and the error bars denote the 16th--84th percentile range, illustrating the finite vertical extent sampled by each line. These heights are used here primarily as approximate proxies for relative ordering rather than as precise geometrical heights. Most heights are negative because of the Wilson depression.
}
\label{fig:context_height}
\end{figure*}

Figure~\ref{fig:context_height} summarises the observational context and the selected lines. The slit-jaw image (left panel) shows the active-region context and the raster field of view, which includes the analysed pore. The stacked line-core raster images in the middle illustrate the pore appearance in the eight selected lines arranged in formation-height order, while the formation-height panel shows the adopted effective heights of the same lines. The error bars indicate the corresponding 16th--84th percentile ranges of the contribution functions and thus the finite vertical extent sampled by each spectral line. The mostly negative height values reflect that the reference zero point is defined relative to the quiet-Sun $\tau_{500\,nm}=1$ surface, so lines formed within the Wilson-depressed pore atmosphere can appear at negative geometrical heights on this scale \citep[e.g.,][]{2003A&ARv..11..153S}.

\section{Analysis and Results}
\label{sec:analysis_results}

All wave analyses carried out in this work were performed with routines implemented in the WaLSAtools\footnote{\url{https://github.com/WaLSAteam/WaLSAtools}} library \citep{2025NRvMP...5...21J, walsatools..2025...17569951}. All signals entering the wave analysis were first linearly detrended and apodised with a Tukey window using a 10\% taper. This preprocessing was applied consistently throughout the temporal and cross-line wavelet analyses.

\subsection{Pore boundary extraction from line-core intensity rasters}
For each of the eight selected spectral lines, we constructed a time series of two-dimensional raster images at the chosen line-core wavelength position and tracked the pore boundary in those images. As noted in Section~\ref{sec:data}, the aim of these rasters is to follow the boundary morphology of the pore rather than to measure line-centre shifts themselves. Boundary identification was therefore performed directly on the intensity images at fixed wavelength positions. The adopted contour level for each line was chosen to isolate the dark pore core while avoiding nearby intergranular lanes and transient brightenings. To improve robustness against small-scale noise and pixel-to-pixel fluctuations, the image sequences were spatially upsampled, and the pore boundary in each frame was determined using an intensity-contour approach with temporal continuity imposed so that the identified contour evolved smoothly in time and did not jump spuriously between unrelated intensity features. The resulting contours were then mapped onto a common centred reference frame, allowing the boundary evolution to be analysed consistently in time for each spectral line and providing a stable basis for comparing azimuthal boundary-shape perturbations across the selected lines.

This boundary-based approach follows the spirit of our earlier pore analysis using SO/PHI-HRT data, where oscillations of the pore boundary were used to isolate the dominant MHD eigenmodes \citep{2024A&A...688A...2J}. In the present case, however, the analysis is extended from a single photospheric layer to a set of near-UV spectral lines spanning different formation heights, for which the centred radial representation provides a consistent basis for cross-line comparison of boundary-shape perturbations. For each line, the final product of this step is a time-dependent pore boundary expressed as a radial displacement $r(\theta,t)$ around a common centre, where $\theta$ is the azimuthal angle along the pore perimeter and $t$ denotes time. This centred radial representation is required in order to compare the boundary as a function of azimuth in a consistent way between time steps and between spectral lines. Without such centring, variations in the absolute position of the pore would be mixed with genuine boundary-shape perturbations in $r(\theta,t)$, making the cross-line comparison less clean. The present formulation therefore isolates coherent boundary-shape perturbations, while a bulk transverse displacement of the pore as a whole is less naturally captured.

\subsection{Proper orthogonal decomposition of the boundary oscillations}
To isolate the dominant coherent boundary perturbations, we applied POD to this time-dependent boundary representation for each selected spectral line. The radial displacement $r(\theta,t)$ was first resampled onto a uniform angular grid and, after subtracting the temporal mean boundary shape, the resulting fluctuations were decomposed into orthogonal spatial eigenfunctions and their associated temporal coefficients according to
\begin{equation}
r(\theta,t) = \overline{r}(\theta) + \sum_{k=1}^{N} a_k(t)\,\phi_k(\theta),
\end{equation}
where $\overline{r}(\theta)$ is the mean boundary shape, $\phi_k(\theta)$ are the POD eigenfunctions, $a_k(t)$ are the corresponding temporal coefficients, $k$ labels the POD mode, and $N$ is the total number of retained POD modes in the decomposition. Here, $k$ denotes only the POD rank index and is distinct from the azimuthal wave number $m$, which is assigned later from the spatial morphology of the recovered eigenfunctions. In matrix form, this decomposition is obtained from the covariance matrix of the mean-subtracted boundary fluctuations. Its eigenvalues, $\lambda_k$, quantify how much of the total fluctuation power is captured by each mode. Throughout this Letter we therefore express the relative importance of each POD mode through the normalised eigenvalue fractions,
\begin{equation}
f_k = \frac{\lambda_k}{\sum_{i=1}^{N}\lambda_i}.
\end{equation}
These fractions are quoted as percentages and describe the proportion of the total boundary variance captured by each mode. We adopt this formulation because it remains directly linked to the POD eigenvalue problem while retaining the more intuitive interpretation in terms of variance contribution.

In this framework, the POD eigenfunctions describe the dominant azimuthal boundary perturbations, while the temporal coefficients quantify how strongly each perturbation evolves in time. POD is therefore particularly well suited to identifying MHD boundary modes, since different azimuthal wave numbers produce distinct spatial morphologies around the pore perimeter when compared with the expectations for a cylindrical magnetic flux tube. In what follows, we focus on the two dominant recovered POD modes, i.e. the first two POD components in rank order, which together account for the overwhelming majority of the total normalised eigenvalue fraction in all selected lines and are the most consistently recovered components in the present centred boundary representation.

\subsection{Identified boundary modes across the selected lines}

\begin{figure*}
\centering
\includegraphics[width=\textwidth]{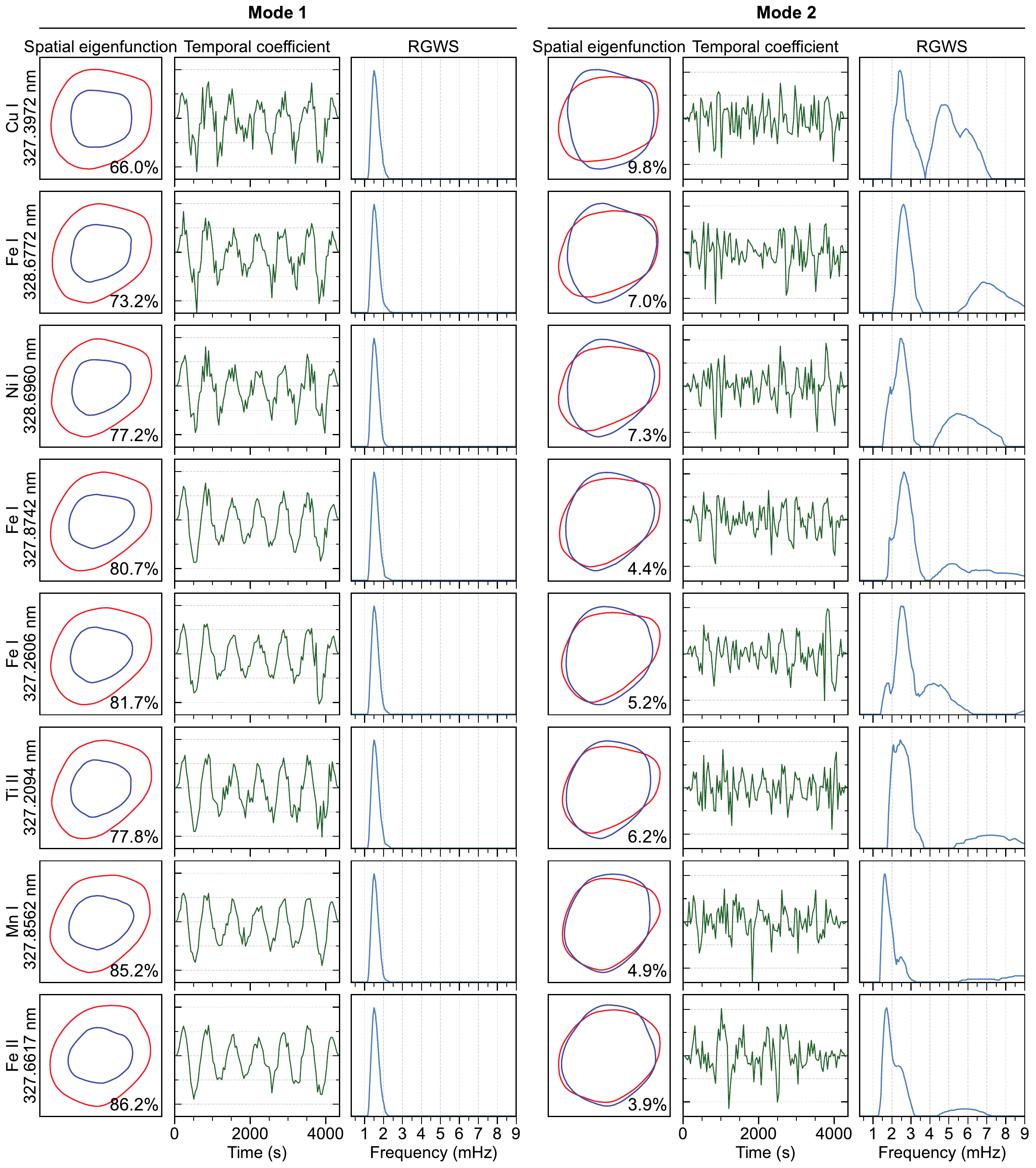}
\caption{Two dominant recovered POD modes identified from the pore-boundary oscillations in the eight selected spectral lines, ordered by increasing formation height from bottom to top. For each line, the first three columns show the first recovered POD mode and the last three columns the second recovered POD mode. Within each group, the panels show the spatial eigenfunction, the corresponding temporal coefficient, and the refined global wavelet spectrum (RGWS). The temporal-coefficient panels are plotted on the same vertical scale within each mode to allow a direct comparison of amplitudes between spectral lines. The percentages quoted in the spatial panels indicate the normalised eigenvalue fraction of each mode, that is, the corresponding eigenvalue divided by the sum of all eigenvalues. The first mode is consistently sausage-like and dominates the total normalised eigenvalue fraction, while the second mode is consistently fluting-like with $m=2$ and contributes a smaller but still coherent fraction. The temporal evolution of the spatial eigenfunctions for the two identified modes is available as an \href{https://walsa.team/sunrise/figure2_animation}{online animation of the figure}.
}
\label{fig:pod_modes}
\end{figure*}

Figure~\ref{fig:pod_modes} presents, for each of the eight selected spectral lines, the spatial form of the two dominant recovered POD modes together with their temporal coefficients and corresponding refined global wavelet spectra (RGWS). The RGWS is defined as the time-integrated Morlet wavelet power as a function of frequency, retaining only frequencies where the local power exceeds the 95\% confidence level and excluding regions affected by the cone of influence (COI) \citep{2025NRvMP...5...21J}. In all cases, the first mode shows the characteristic axisymmetric contraction--expansion pattern expected for a sausage ($m=0$) mode, while the second mode exhibits the alternating azimuthal structure expected for a fluting mode with $m=2$. Because the present centred boundary representation is designed to emphasise boundary-shape perturbations rather than bulk transverse displacement of the pore as a whole, no comparably robust kink-like ($m=1$) component is recovered in the current decomposition.

The dominant mode identified in every selected spectral line is thus an axisymmetric sausage mode. Its temporal coefficient shows a clear narrow-band oscillatory signal whose amplitude remains broadly comparable across the line ensemble, and its RGWS is remarkably similar from line to line, with a prominent concentration of power in the 1--2\,mHz range. The strong consistency of both the spatial morphology and the dominant temporal power indicates that the same physical mode is being sampled across all selected lines.

The second POD mode is consistently fluting-like, with an azimuthal structure corresponding to $m=2$. Although it contributes substantially less of the total normalised eigenvalue fraction than the sausage mode, it is nevertheless clearly recovered in all eight lines. Its temporal coefficients also show coherent oscillatory behaviour, but with broader and somewhat more line-dependent power spectra than those of the first mode. In most cases, the dominant power of the fluting mode lies in the 2--3.5\,mHz range, while secondary power enhancements at higher frequencies, extending up to about 9\,mHz, become more apparent in the higher-forming lines.

The relative importance of the two modes is highly systematic across the selected lines. The first POD mode typically contributes about 66--86\% of the total normalised eigenvalue fraction, whereas the second contributes about 4--10\%. Overall, the fractional contribution of the sausage mode tends to decrease with increasing formation height, while that of the fluting mode tends to increase. At the same time, the positive and negative extrema of the fluting perturbation become more clearly separated in the higher-forming lines, suggesting that the $m=2$ boundary distortion becomes morphologically more pronounced with height, even though the amplitudes of the temporal coefficients remain broadly comparable across the selected lines.

Two points are particularly noteworthy. First, the persistence of the same two leading morphologies across the entire line ensemble demonstrates that these are not line-specific artefacts of the boundary extraction, but robust physical modes of the pore. Second, the dominance of the sausage mode in the normalised eigenvalue budget is itself a clear result: the boundary dynamics are overwhelmingly governed by an $m=0$ oscillation, while the fluting mode appears as a weaker but still coherent higher-order perturbation. This behaviour is broadly consistent with the earlier single-layer identification of sausage and fluting modes in a pore \citep{2024A&A...688A...2J}, but is established here simultaneously across multiple near-UV spectral lines and hence across multiple sampled heights.

\subsection{Cross-line coherence and phase behaviour}
To investigate how these identified modes behave with height, we compared the temporal coefficients of corresponding POD modes between the selected spectral lines using wavelet coherence analysis. For each mode, the temporal coefficient of the lowest-forming selected line, Fe\,{\sc ii} 327.6617\,nm, was adopted as the reference signal. Each of the remaining seven coefficients was then compared with this reference using the Morlet-wavelet framework in WaLSAtools. From these cross-wavelet calculations, we extracted the wavelet coherence and phase angle as functions of time and frequency.

Phase information was retained only in regions where the coherence was both statistically significant at the 95\% confidence level and located outside the COI. This restriction ensures that the inferred phase angles are based only on the coherent part of the signal and are not dominated by edge effects or non-significant fluctuations. For each POD mode, we then identified the frequency interval that provided the most robust common coherence across the selected lines and computed the corresponding consensus phase statistics.

\begin{figure}
\centering
\includegraphics[width=\columnwidth]{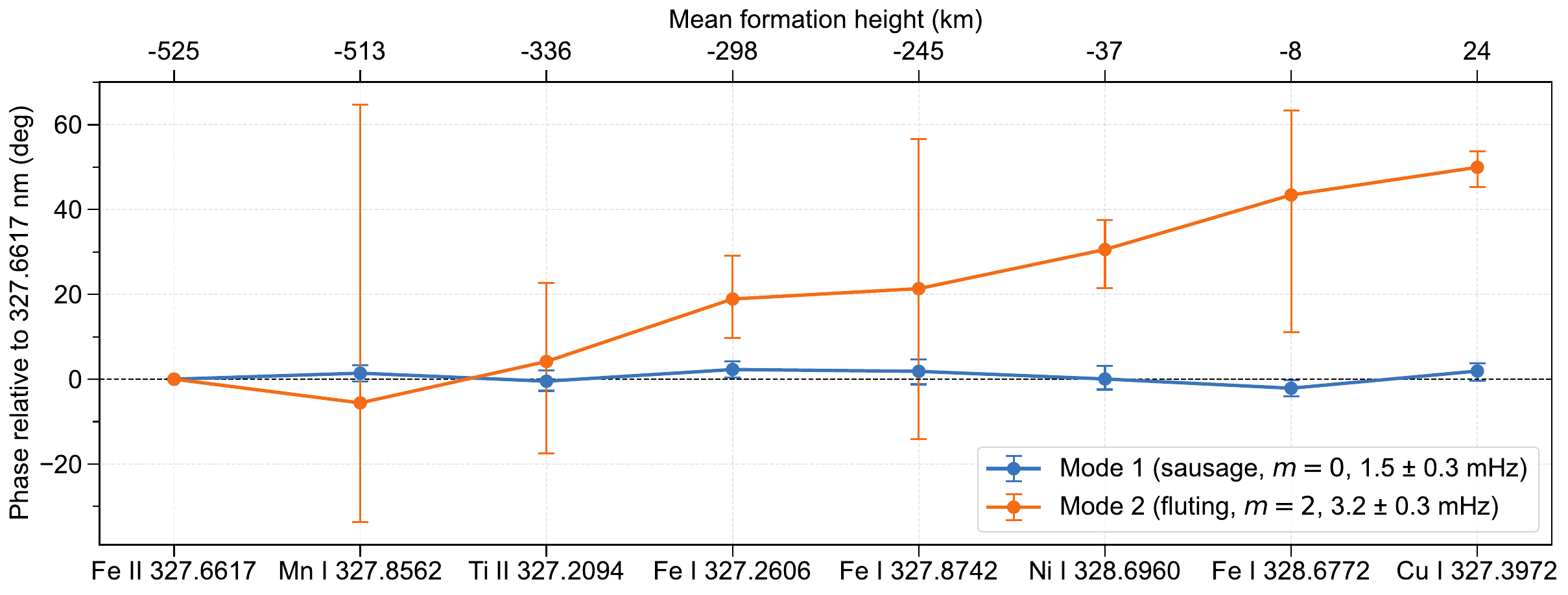}
\caption{Consensus phase behaviour of the two dominant recovered POD modes across the selected spectral lines. The phase is measured relative to the lowest-forming selected line, Fe\,{\sc ii} 327.6617\,nm, and is shown for the frequency bands in which the common coherence is strongest: $1.5\pm0.3$\,mHz for the sausage mode and $3.2\pm0.3$\,mHz for the fluting mode. Only phase values from regions outside the cone of influence and above the 95\% coherence significance threshold are retained. The plotted phase at each spectral line is the coherence-weighted mean phase, and the error bars denote the corresponding 16th--84th percentile-based phase spread. The lower horizontal axis lists the spectral lines, while the upper axis gives their adopted mean formation heights. The sausage mode remains close to zero phase difference across the sampled heights, consistent with standing or near-standing behaviour, whereas the fluting mode shows a modest but systematic increase in phase with height, indicative of an upward-propagating component.
}
\label{fig:phase_height}
\end{figure}

For the first POD mode, the strongest common coherent signal lies in the 1.2--1.8\,mHz range. Across this band, the mean phase differences relative to the reference line remain very close to zero for all selected lines, with only small deviations within the phase spreads. Figure~\ref{fig:phase_height} shows that the phase behaviour of this mode is therefore essentially flat with height. This result is consistent with a standing, or at least near-standing, sausage oscillation sampled through the selected atmospheric layers.

For the second POD mode, the strongest common coherent signal is found in the 2.9--3.5\,mHz range. For this mode, the mean phase does not remain constant. Instead, it shows a modest but systematic increase across the line ensemble, rising from values close to zero in the lowest-forming line to roughly $50^{\circ}$ in the highest-forming one (Figure~\ref{fig:phase_height}). Although the increase is not perfectly linear from line to line, the overall trend is clear and indicates an upward-propagating component for the fluting mode. Interpreted literally, the approximately $50^{\circ}$ phase change across the full adopted height range at $\sim$3.2~mHz would correspond to an apparent phase speed of order 10~km{\,}s$^{-1}$. However, any corresponding range of values would be dominated by the substantial uncertainty in the effective formation heights and the broad vertical extent of the contribution functions. We therefore do not treat this value as a seismological measurement. Instead, we only provide this estimate to show that the observed phase progression has a physically plausible scale.

Taken together, these results provide a simple but physically informative picture of the pore boundary dynamics. The dominant boundary oscillation is a standing or near-standing sausage mode that remains phase-aligned across the sampled heights, while the weaker fluting mode exhibits a measurable upward phase progression. The combination of POD-based mode identification and coherence-restricted phase analysis therefore allows the vertical behaviour of the leading pore eigenmodes to be diagnosed directly from the multi-line near-UV observations.

\section{Discussion and Conclusions}
\label{sec:discussion_conclusions}

In this Letter, we used seeing-free near-UV observations from {\sc Sunrise-iii}/SUSI to examine the boundary oscillations of a small solar pore across eight selected spectral lines in the 327--329\,nm window. By applying POD to the boundary oscillations and then comparing the temporal coefficients through wavelet coherence and phase analysis, we obtained a direct multi-height view of the dominant pore boundary modes and their propagation behaviour.

In all eight selected spectral lines, the first POD mode is consistently identified as an axisymmetric sausage mode and carries the dominant normalised eigenvalue fraction, typically about 66--86\%. The second POD mode is consistently fluting-like, with azimuthal wave number $m=2$, and contributes about 4--10\%. The phase analysis then shows that these two modes differ not only in morphology but also in their vertical behaviour: the sausage mode remains close to zero phase difference across the selected lines, consistent with standing or near-standing behaviour, whereas the fluting mode exhibits a modest but systematic phase increase with height, indicating an upward-propagating component.

These results extend earlier pore-wave studies in two main respects. First, previous investigations had already established the presence of sausage oscillations in pores, including both standing and upwardly propagating cases \citep[e.g.,][]{2014A&A...563A..12D, 2015A&A...579A..73M, 2022ApJ...938..143G}. Second, an earlier seeing-free SO/PHI-HRT study showed that sausage, kink, and fluting modes can coexist in a pore, but only at a single sampled photospheric layer \citep{2024A&A...688A...2J}. The present Letter advances that picture by demonstrating that the dominant sausage and fluting modes can be identified consistently across multiple near-UV spectral lines sampling heights differing by roughly 500\,km from each other, within a centred radial boundary representation chosen to enable a consistent cross-line comparison of boundary-shape perturbations. In that sense, this is the first multi-height observational characterisation of these two pore oscillations, particularly with such fine sampling through the photospheric and low-chromospheric regime. In this formulation, the sausage- and fluting-like components are the most consistently recovered modes across the selected line ensemble. At the same time, it is less naturally sensitive to a bulk transverse displacement of the pore as a whole, and therefore does not favour the recovery of a kink-like component in the same way as a decomposition that retains the absolute pore position.

This distinction is also the main difference from \citet{2024A&A...688A...2J}. In that earlier single-layer study, the analysis framework was able to recover sausage, kink, and fluting components in the boundary oscillations. Here, by contrast, the aim is not to compile the broadest possible mode inventory at one height, but to compare the same pore boundary consistently across a sequence of spectral lines sampling different atmospheric layers. For that purpose, the common centred radial representation provides a consistent azimuthal description of the same pore boundary at all times and in all selected lines. Without such centring, variations in the absolute position of the pore would be mixed with genuine boundary-shape perturbations in the cross-line comparison, making the identification of the coherent sausage- and fluting-like components less clean. A dedicated treatment of kink-like motions, which would require retaining and analysing the transverse displacement of the pore as a whole, is beyond the scope of the present Letter and is deferred to future work. We therefore do not interpret the absence of a comparably robust kink-like mode in the present decomposition as evidence that such motions are absent from the pore.

The relative contribution of the sausage and fluting modes also changes systematically across the selected line ensemble, with the sausage fraction tending to decrease and the fluting fraction tending to increase with height. Because these normalised eigenvalue fractions can be influenced not only by the underlying wave behaviour, but also by line-dependent formation properties, boundary-identification sensitivity, and the centred radial representation adopted here, we do not attempt a more detailed physical interpretation of this trend in the present Letter. Disentangling these effects will require dedicated forward modelling and MHD simulations, which are beyond the scope of the present Letter.

The phase behaviour is especially informative. The nearly zero phase differences of the sausage mode across the selected lines are consistent with standing or near-standing behaviour, in clear contrast to the $m=2$ fluting mode at higher frequency, which exhibits a systematic phase progression with height. A purely evanescent disturbance could also, in principle, produce very small phase differences over the sampled height range. In the classical picture, an evanescent wave would be associated with an exponentially decaying amplitude with height, while a standing disturbance sampled over only a relatively small fraction of its full wavelength would be expected to show an approximately constant amplitude over the limited height range considered here. Our observed amplitudes are essentially constant across the selected lines, which is most naturally consistent with a standing or near-standing interpretation. We acknowledge that in a stratified atmosphere the amplitude behaviour of an evanescent disturbance can be more subtle than the strict exponential decay of the uniform-medium case. A weakly evanescent disturbance therefore cannot be fully excluded from the present analysis. The apparent amplitudes are additionally influenced by line-dependent formation characteristics, the finite vertical extent of the contribution functions, radiative-transfer response, line-core contrast, and differences in boundary-identification sensitivity between spectral lines, further limiting their use as a quantitative discriminator. We therefore adopt a standing or near-standing description of the sausage mode as the most natural interpretation of the observations, while noting that a weakly evanescent disturbance is not excluded. A more definitive discrimination between these scenarios will require dedicated forward MHD modelling together with a full-Stokes polarimetric analysis of the pore, and we defer that to future work.

For the fluting mode, the interpretation is more straightforward. Its systematic phase increase with height is itself the clearest signature of an upward-propagating component and is not consistent with a purely standing or purely evanescent disturbance. The absence of a clear amplitude decrease with height does not provide a similarly strong additional constraint here; rather, the phase behaviour is the robust result. At the same time, the spatial appearance of the fluting mode becomes more clearly expressed in the higher-forming lines, and its power spectra show a richer high-frequency content there, while the common dominant power remains concentrated near 2--3.5\,mHz throughout the line ensemble. This suggests that the $m=2$ perturbation not only propagates upward, but also becomes more morphologically pronounced and spectrally structured with height.

Because of the substantial uncertainty in the exact formation heights and the finite vertical widths of the contribution functions, we deliberately avoid a strong quantitative phase-speed claim in this early-science study. If the observed phase progression of the fluting mode were interpreted literally in terms of upward propagation through the adopted height ordering, it would imply only an order-of-magnitude phase speed, and such an estimate would remain dominated by the present height uncertainties. We therefore regard the qualitative distinction between the two modes as the more robust result: the dominant sausage mode is phase-aligned across the sampled layers, whereas the fluting mode is not.

It is also important to emphasise the deliberate scope of the present analysis. This study is based exclusively on SUSI Stokes~$I$ data and was designed as an early-science investigation of pore boundary oscillations using the currently available products. No polarimetric information was used here, because the corresponding full-Stokes profiles were not yet ready at this stage of the {\sc Sunrise-iii}/SUSI programme. The present results should therefore be viewed as a focused first step based on intensity-defined boundary motions alone. Even within that restricted scope, however, the near-UV SUSI observations already reveal a remarkably coherent and physically interpretable mode structure.

The broader importance of this work lies in what the SUSI data now make possible. The line-rich near-UV window allows the same magnetic structure to be examined simultaneously through several closely spaced spectral lines, providing leverage that is difficult to obtain from one- or two-line studies alone. In the present pore, this makes it possible not only to identify the dominant boundary eigenmodes, but also to trace how their phase behaviour changes across the sampled layers. Together with Papers~I and II, and with other recent {\sc Sunrise-iii}/SUSI wave studies using complementary many-line diagnostics \citep{Lagg+2026}, the present Letter therefore strengthens the emerging picture that {\sc Sunrise-iii}/SUSI can resolve wave behaviour in active-region magnetic structures in a genuinely multi-line and multi-height sense.

Several natural extensions follow from this work. Future studies can combine the present boundary analysis with complementary inversions and, once the relevant polarimetric products become available, relate the identified modes more directly to the local thermal and magnetic structure of the pore. The present Letter also concentrates on surface boundary oscillations; a logical next step is to investigate the oscillatory response of the pore interior itself, where body-type wave behaviour and internal structuring may also be present. In addition, combining SUSI with the other {\sc Sunrise-iii} instruments, SCIP \citep{2020SPIE11447E..0YK} and TuMag \citep{2025SoPh..300..148D}, will allow the vertical behaviour of these modes to be followed over a larger atmospheric range. Finally, other SUSI spectral windows, including those sampling higher chromospheric layers, offer the prospect of tracing whether the identified boundary modes persist, transform, or dissipate further upward. More generally, the same approach could in future be extended to observations sampling the chromosphere, transition region, and low corona, allowing coherent wave modes to be traced over a much larger height range. This would also help to test how MHD modes evolve as magnetic flux tubes expand and change cross-sectional shape with height.

In summary, we have presented the first multi-height observational characterisation of sausage and fluting modes in a solar pore, as recovered from a centred radial analysis of the pore-boundary oscillations. The leading mode is an $m = 0$ sausage oscillation that dominates the boundary-variance budget and remains nearly phase-aligned across the sampled response heights, while the second mode is an $m = 2$ fluting oscillation with a measurable upward phase progression. These results provide the first multi-height view of sausage and fluting modes in a pore and demonstrate the considerable potential of the {\sc Sunrise-iii} near-UV data for future studies of MHD wave structuring in small-scale solar magnetic elements.

\begin{acknowledgments}
SJ and DBJ acknowledge support from the UK Science and Technology Facilities Council (STFC) through consolidated grants ST/T00021X/1 and ST/X000923/1. SJ also received support from the Rosseland Centre for Solar Physics (RoCS), University of Oslo, Norway. DBJ further acknowledges funding from the Leverhulme Trust (Research Project Grant RPG-2019-371) and from the UK Space Agency via the National Space Technology Programme (grant SSc-009).
LS acknowledges support from STFC grant ST/X001008/1. SSAS, VF and GV are grateful to the STFC grants ST/V000977/1, ST/Y001532/1 and The Royal Society, International Exchanges Scheme 2023 Cost Share (NSTC), collaboration with Taiwan (IEC/R3/233017). VF is grateful to the STFC grant UKRI1165. SSAS and VF would like to thank the International Space Science Institute (ISSI Team 607). 
We wish to acknowledge scientific discussions with the Waves in the Lower Solar Atmosphere (WaLSA; \href{https://WaLSA.team}{www.WaLSA.team}) team, which has been supported by the Research Council of Norway (project no. 262622), The Royal Society (award no. Hooke18b/SCTM; \citealt{2021RSPTA.37900169J}), and the International Space Science Institute (ISSI Team 502).
\textsc{Sunrise iii} is supported by funding from the Max-Planck-F\"orderstiftung (Max Planck Foundation), NASA under Grants \#80NSSC18K0934 and \#80NSSC24M0024 (``Heliophysics Low Cost Access to Space'' program), and the ISAS/JAXA Small Mission-of-Opportunity program and JSPS KAKENHI Grant Numbers JP18H05234 and JP23K25916. This research has received financial support from the European Union's Horizon 2020 research and innovation programme under grant agreement No.~824135 (SOLARNET) and No.~101097844 (WINSUN) from the European Research Council (ERC). It has also been funded by the Deutsches Zentrum f\"ur Luft- und Raumfahrt e.V.\ (DLR, grant no.~50~OO~1608). The Spanish contributions have been funded by the Spanish MCIN/AEI under projects RTI2018-096886-B-C5 and PID2021-125325OB-C5, and from ``Center of Excellence Severo Ochoa'' awards to IAA-CSIC (SEV-2017-0709, CEX2021-001131-S), all co-funded by European REDEF funds, ``A way of making Europe''.
\end{acknowledgments}

\bibliography{article}{}
\bibliographystyle{aasjournalv7}

\end{document}